\documentclass[aps,prd,12pt,amssymb]{revtex4}

\begin{document}

\baselineskip=0.60cm

\newcommand{\ini}{\begin{equation}}
\newcommand{\fin}{\end{equation}}
\newcommand{\inir}{\begin{eqnarray}}
\newcommand{\finr}{\end{eqnarray}}
\newcommand{\inif}{\begin{figure}}
\newcommand{\finf}{\end{figure}}
\newcommand{\bc}{\begin{center}}
\newcommand{\ec}{\end{center}}

\def\ol{\overline}
\def\pa{\partial}
\def\ra{\rightarrow}
\def\ts{\times}
\def\df{\dotfill}
\def\bs{\backslash}
\def\dg{\dagger}

$~$

\hfill DSF-07/2003

\vspace{1 cm}

\title{MINIMAL SEESAW MECHANISM}

\author{D. Falcone}

\affiliation{Dipartimento di Scienze Fisiche,
Universit\`a di Napoli, Via Cintia, Napoli, Italy}

\begin{abstract}
\vspace{1cm}
\noindent
In the framework of the seesaw mechanism, and adopting a typical form
for the Dirac neutrino mass matrix, we discuss the impact of
minimal forms of the Majorana neutrino mass matrix. These matrices contain
four or three texture zeros and only two parameters, a scale factor and
a hierarchy parameter. Some forms are not compatible with large lepton
mixing and are ruled out. Moreover, a normal mass hierarchy for neutrinos
is predicted.
\end{abstract}

\maketitle

\newpage

\section{Introduction}

There is now a strong evidence for neutrino oscillations, especially through
the SuperKamiokande, K2K, and SNO, KamLAND experiments \cite{exp}.
Neutrino oscillations are naturally accounted for if neutrinos have small
masses, so that leptons can mix in a similar way as quarks do \cite{mix}.
Moreover, small neutrino masses can be achieved by means of the
seesaw mechanism \cite{ss}.
In this framework, the effective (Majorana) mass matrix of neutrinos $M_L$
is related to the Dirac neutrino mass matrix $M_{\nu}$ and the heavy Majorana
neutrino mass matrix $M_R$ by the relation
\ini
M_L \simeq M_{\nu} M_R^{-1} M_{\nu}^T.
\fin
As a matter of fact, the seesaw formula (1) is valid at the high $M_R$ scale,
and therefore one should determine both $M_{\nu}$ and $M_L$ at that scale,
in order to find a consistent model. The effective matrix $M_L$ is partially
described at the low scale through the analysis of several experiments
\cite{fs1}.
On the other hand, the Dirac matrix $M_{\nu}$ is based on theoretical hints.
Both have to be renormalized to the $M_R$ scale. Then the problem is to find
models for $M_{\nu}$ and $M_R$ which reproduce the phenomenological forms
of $M_L$ according to the master relation (1). Such a problem has been addressed
in many papers (see, for instance, the review \cite{bd}).
In the present article we consider a structure for $M_{\nu}$
which is usually adopted for charged fermion mass matrices, and
minimal models for $M_R$. We select minimal forms which are compatible with
phenomenology.

In section II we discuss the effective neutrino mass matrix. In section III
we describe the Dirac and Majorana mass matrices of our minimal framework.
Then, in section IV, the seesaw formula is applied and the resulting neutrino
mass matrix is compared to phenomenology.
A brief discussion is finally proposed.

\section{The effective neutrino mass matrix}

Experimental informations on neutrino oscillations imply that the lepton
mixing matrix is given by
\ini
U \simeq \left( \begin{array}{ccc}
\frac{\sqrt2}{\sqrt3} & \frac{1}{\sqrt3} &
\epsilon \text{e}^{-\text{i} \delta} \\
-\frac{1}{\sqrt6}-\frac{1}{\sqrt3}\epsilon \text{e}^{\text{i} \delta} &
\frac{1}{\sqrt3}-\frac{1}{\sqrt6}\epsilon \text{e}^{\text{i} \delta} &
\frac{1}{\sqrt2} \\
\frac{1}{\sqrt6}-\frac{1}{\sqrt3}\epsilon \text{e}^{\text{i} \delta} &
-\frac{1}{\sqrt3}-\frac{1}{\sqrt6}\epsilon \text{e}^{\text{i} \delta} &
\frac{1}{\sqrt2}
\end{array} \right)
\text{diag}(\text{e}^{\text{i} \varphi_1/2},\text{e}^{\text{i} \varphi_2/2},1)
\fin
in the standard parametrization, where $\epsilon<0.16$, $0<\delta<2\pi$,
$0 < \varphi_1,\varphi_2 < 2\pi$ (see, for example, Ref.\cite{tribi}).
Moreover, neutrino oscillations determine the following square mass
differences,
\ini
\Delta m^2_{32}=m_3^2-m_2^2 \simeq 3 \cdot 10^{-3} \text{eV}^2,
\fin
\ini
\Delta m^2_{21}=m_2^2-m_1^2 \simeq 7 \cdot 10^{-5} \text{eV}^2,
\fin
where $m_1,m_2,m_3$ are the effective neutrino masses. Then,
in the basis where the charged lepton mass matrix is diagonal,
$M_L$ is obtained by means of the transformation
\ini
M_L =U^* D_L U^{\dg},
\fin
with $D_L=\text{diag}(m_1,m_2,m_3)$.
Neglecting $\epsilon$ in $U$, except for $U_{e3}$, the calculation leads to
$$
M_{ee} \simeq \epsilon^2 m_3+\frac{m_2}{3}+2\frac{m_1}{3}
$$
$$
M_{e \mu} \simeq \epsilon \frac{m_3}{\sqrt2}+\frac{m_2}{3}-\frac{m_1}{3}
$$
$$
M_{e \tau} \simeq \epsilon \frac{m_3}{\sqrt2}-\frac{m_2}{3}+\frac{m_1}{3}
$$
$$
M_{\mu \mu} \simeq M_{\tau \tau} \simeq \frac{m_3}{2}+\frac{m_2}{3}+\frac{m_1}{6}
$$
$$
M_{\mu \tau} \simeq \frac{m_3}{2}-\frac{m_2}{3}-\frac{m_1}{6}
$$
not writing phases, which can be inserted by
$\epsilon \ra \epsilon \text{e}^{\text{i} \delta}$,
$m_1 \ra m_1 \text{e}^{\text{i} \varphi_1}$,
$m_2 \ra m_2 \text{e}^{\text{i} \varphi_2}$.

Since $\Delta m^2_{21} \ll \Delta m^2_{32}$, we have two main mass spectra for
light neutrinos, the normal hierarchy $m_1 < m_2 \ll m_3$, and the inverse
hierarchy $m_1 \simeq m_2 \gg m_3$. For the normal hierarchy the dominant
elements are given by
\ini
M_L \sim \left( \begin{array}{ccc}
0 & 0 & 0 \\ 0 & \frac{1}{2} & \frac{1}{2} \\ 0 & \frac{1}{2} & \frac{1}{2} 
\end{array} \right) m_3,
\fin
with $m_3 \simeq \sqrt{\Delta m^2_{32}}$,
and for the inverse hierarchy they are given by
\ini
M_L \sim \left( \begin{array}{ccc}
1 & 0 & 0 \\ 0 & \frac{1}{2} & -\frac{1}{2} \\ 0 & -\frac{1}{2} & \frac{1}{2} 
\end{array} \right) m_{1,2},
\fin
with $m_{1,2} \simeq \sqrt{\Delta m^2_{32}}$. 
Both contain a democratic $\mu \tau$ block, due to near maximal $U_{\mu 3}$.
The difference stands in the element $ee$, which is suppressed in (6)
but dominant in (7).
Now, according to Ref.\cite{fs2}, the general structure of $M_L$ is not
changed by renormalization. Therefore, we can take
matrices (6) and (7) as simple forms at the high scale, the zero elements
meaning suppressed with respect to dominant elements. They correspond to
distinct predictions for the double beta decay parameter $M_{ee}$, since
for the normal hierarchy we get
$M_{ee} \sim \sqrt{\Delta m^2_{21}}$, while
for the inverse hierarchy we have
$M_{ee} \sim \sqrt{\Delta m^2_{32}}$.

\section{Dirac and Majorana mass matrices}

In order to apply the seesaw formula, we need the expression of the
Dirac and Majorana mass matrices.
We take a typical form for the three mass matrices of charged fermions \cite{cf}:
\ini
M_d \simeq \left( \begin{array}{ccc}
0 & \sqrt{m_d m_s} & 0 \\
\sqrt{m_d m_s} & m_s & \sqrt{m_d m_b} \\
0 & \sqrt{m_d m_b} & m_b,
\end{array} \right)
\fin
\ini
M_u \simeq \left( \begin{array}{ccc}
0 & \sqrt{m_u m_c} & 0 \\
\sqrt{m_u m_c} & m_c & \sqrt{m_u m_t} \\
0 & \sqrt{m_u m_t} & m_t 
\end{array} \right)
\fin
for down and up quarks, and
\ini
M_e \simeq \left( \begin{array}{ccc}
0 & \sqrt{m_e m_{\mu}} & 0 \\
\sqrt{m_e m_{\mu}} & m_{\mu} & \sqrt{m_e m_{\tau}} \\
0 & \sqrt{m_e m_{\tau}} & m_{\tau} 
\end{array} \right)
\fin
for charged leptons.
Then, since $M_e \sim M_d$, a natural choice is also $M_{\nu} \sim M_u$.
In fact, we have \cite{br} $m_d/m_s \sim m_s/m_b \sim \lambda^2$,
$m_e/m_{\mu} \sim m_{\mu}/m_{\tau} \sim \lambda^2$, and
$m_u/m_c \sim m_c/m_t \sim \lambda^4$, where $\lambda=0.22$ is the Cabibbo
parameter. The renormalization of quark mass matrices does not affect their
expression in terms of powers of $\lambda$ \cite{bbo}.
Therefore, for the Dirac neutrino mass matrix we take
\ini
M_{\nu} \simeq \left( \begin{array}{ccc}
0 & a & 0 \\ a & b & c \\ 0 & c & 1
\end{array} \right) m_t,
\fin
with $a \ll b \sim c \ll 1$.
As order in $\lambda$ we have $a \sim \lambda^6$, $c \sim \lambda^4$.
Expressions (8) and (9) lead to small quark mixings, while lepton mixings
$U_{e2}$ and $U_{\mu 3}$ are large. The matrix $M_R$ should produce,
through the seesaw formula, large lepton mixings \cite{smir}. 

For this Majorana mass matrix we consider minimal forms. These include matrices
with four texture zeros:
\ini
M_R = \left( \begin{array}{ccc}
0 & A & 0 \\ A & 0 & 0 \\ 0 & 0 & B
\end{array} \right) m_R,
\fin
\ini
M_R = \left( \begin{array}{ccc}
A & 0 & 0 \\ 0 & 0 & B \\ 0 & B & 0
\end{array} \right) m_R,
\fin
\ini
M_R = \left( \begin{array}{ccc}
0 & 0 & A \\ 0 & B & 0 \\ A & 0 & 0
\end{array} \right) m_R,
\fin
and matrices with three texture zeros, that is the diagonal form
\ini
M_R = \left( \begin{array}{ccc}
A & 0 & 0 \\ 0 & B & 0 \\ 0 & 0 & C
\end{array} \right) m_R,
\fin
and the Zee-like form \cite{stech}
\ini
M_R = \left( \begin{array}{ccc}
0 & A & B \\ A & 0 & C \\ B & C & 0
\end{array} \right) m_R.
\fin
Parameters $A,~B,~C$ can take values $1$ and $r < 1$. Therefore, in such
matrices there is a scale factor, related to $m_R$, and possibly one hierarchy
parameter $r$.

\section{Seesaw mechanism}

In this section we calculate the effective neutrino mass matrix by means of
the seesaw formula (1) on mass matrices discussed in the previous section.
Then we look for structure of the kind (6) and (7). We exclude possible
cancellations during our analysis.

\subsection{Four texture zeros}

Matrix (12) leads to 
\ini
M_L \simeq \left( \begin{array}{ccc}
0 & \frac{a^2}{A} & 0 \\
\frac{a^2}{A} & \frac{2ab}{A}+\frac{c^2}{B} & \frac{ac}{A}+\frac{c}{B} \\
0 & \frac{ac}{A}+\frac{c}{B} & \frac{1}{B}
\end{array} \right) \frac{m_t^2}{m_R}.
\fin
Condition $M_{\mu \tau} \sim M_{\tau \tau}$ gives $A/B \sim ac$. Then
$A=r \sim ac$ and $B=1$. Condition $M_{\mu \mu} \sim M_{\tau \tau}$ is
satisfied as a consequence. See also Ref.\cite{bo} for a discussion on
this structure.

Matrix (13) leads to
\ini
M_L \simeq \left( \begin{array}{ccc}
0 & \frac{ac}{B} & \frac{a}{B} \\
\frac{ac}{B} & \frac{a^2}{A}+\frac{2bc}{B} & \frac{b}{B}+\frac{c^2}{B} \\
\frac{a}{B} & \frac{b}{B}+\frac{c^2}{B} & \frac{2c}{B}
\end{array} \right) \frac{m_t^2}{m_R}.
\fin
Here condition $M_{\mu \tau} \sim M_{\tau \tau}$ is satisfied while
$M_{\mu \mu} \sim M_{\tau \tau}$ requires $A/B \sim a^2/c$. Hence
$A=r \sim a^2/c$ and $B=1$.

Matrix (14) leads to
\ini
M_L \simeq \left( \begin{array}{ccc}
\frac{a^2}{B} & \frac{ab}{B} & \frac{ac}{B} \\
\frac{ab}{B} & \frac{b^2}{B}+\frac{2ac}{A} & \frac{bc}{B}+\frac{a}{A} \\
\frac{ac}{B} & \frac{bc}{B}+\frac{a}{A} & \frac{c^2}{B}
\end{array} \right) \frac{m_t^2}{m_R},
\fin
so that $M_{\mu \tau} \sim M_{\tau \tau}$ gives $B/A \lesssim c^2/a$.
Then $A=1$ and $B=r \lesssim c^2/a$. The
condition $M_{\mu \mu} \sim M_{\tau \tau}$ is valid as a consequence.

The normal hierarchy is achieved in all cases. However, note the three different
values for the scale $m_R$, that is $m_t^2/m_3$, $cm_t^2/m_3$, $a m_t^2/m_3$,
respectively.

\subsection{Diagonal form}

In this case the effective neutrino mass matrix is given by
\ini
M_L \simeq \left( \begin{array}{ccc}
\frac{a^2}{B} & \frac{ab}{B} & \frac{ac}{B} \\
\frac{ab}{B} & \frac{a^2}{A}+\frac{b^2}{B}+\frac{c^2}{C} &
\frac{bc}{B}+\frac{c}{C} \\
\frac{ac}{B} & \frac{bc}{B}+\frac{c}{C} & \frac{c^2}{B}+\frac{1}{C}
\end{array} \right) \frac{m_t^2}{m_R}.
\fin
Here condition $M_{\mu \tau} \sim M_{\tau \tau}$ gives $B/C \lesssim c^2$.
Then $C=1$ and $B=r \lesssim c^2$. Both $A=r$ and $A=1$ are consistent
with $M_{\mu \mu} \sim M_{\tau \tau}$. The normal hierarchy is obtained.
The scale $m_R$ is given by $m_t^2/m_3$.
Large lepton mixing can indeed be obtained even by means of small mixing
in $M_{\nu}$ and zero mixing in $M_R$ (see Ref.\cite{afm}).
In particular, for $A \simeq B \simeq c^2$ we get
\ini
M_L \simeq \left( \begin{array}{ccc}
k^2 & k & k \\
k & 1 & 1 \\
k & 1 & 1
\end{array} \right) \frac{m_t^2}{m_R},
\fin
with $k=a/c$. This form of $M_L$ has already been proposed several times
\cite{vis}. Moreover, the same form is realized in (19) for $B \simeq c^2/a$,
but with the scale $m_R$ suppressed by the factor $a$ with respect to (20).

\subsection{Zee-like form}

In this case, apart from an overall scale $m_t^2/2m_R$, we get the following
approximate effective matrix
\ini
\left( \begin{array}{ccc}
[-\frac{{a^2}B}{AC}] & [-\frac{abB}{AC}+\frac{a^2}{A}+\frac{ac}{C}] 
& [-\frac{acB}{AC}+\frac{a}{C}] \\
{*} & [-\frac{{a^2}C}{AB}-\frac{{b^2}B}{AC}-\frac{{c^2}A}{BC}+
\frac{2ab}{A}+\frac{2ac}{B}+\frac{2bc}{C}] 
& [-\frac{bcB}{AC}-\frac{cA}{BC}+\frac{ac}{A}+
\frac{a}{B}+\frac{c^2}{C}+\frac{b}{C}] \\
{*} & {*} & [-\frac{{c^2}B}{AC}-\frac{A}{BC}+\frac{2c}{C}]
\end{array} \right).
\fin
In order to have a useful $\mu \tau$ block, the leading terms must be those
with $AC$ in the denominator. Then
the normal hierarchy is achieved for $A=C=r \lesssim c$ and $B=1$.
Here the scale $m_R$ is about $m_t^2/m_3$.

\section{Discussion}

We have studied the seesaw mechanism assuming simple forms of
the fermion mass matrices and in particular minimal forms for the heavy
neutrino mass matrix, which contain four or three texture zeros,
a scale factor and a hierarchy parameter.
Our minimal framework allows only the normal hierarchy
for light neutrinos and not the inverse hierarchy.
Large lepton mixing is achieved by tuning the hierarchy parameter $r$
in the heavy neutrino mass matrix.

There is another possible mass spectrum for neutrinos, the degenerate
spectrum, $m_1 \simeq m_2 \simeq m_3$, which gives the dominant elements
$M_L \sim \text{diag}(1,1,1)m_{1,2,3}$ or
\ini
M_L \sim \left( \begin{array}{ccc}
1 & 0 & 0 \\ 0 & 0 & 1 \\ 0 & 1 & 0 
\end{array} \right) m_{1,2,3}.
\fin
One can easily check that such forms are not reproduced in our minimal
framework. However, both the inverse hierarchy and the degenerate spectrum
can be achieved in some nonminimal models \cite{fal}.
Indeed, generally it is quite hard to yield degeneracy in $M_L$ from
hierarchy in $M_{\nu}$ by means of $M_R$ in the seesaw formula.

The present framework could also be embedded in a unified $SO(10)$ model.
In fact, the relations $M_e \sim M_d$ and $M_{\nu} \sim M_u$ can be the
result of a quark-lepton symmetry, and the high $M_R$ scale can as well
be related to the unification or intermediate breaking scale of the
supersymmetric or nonsupersymmetric model, respectively \cite{falc}.
Then, matrix (13) and especially matrix (14) correspond to the nonsupersymmetric
model, while matrices (12), (15) and possibly (16) correspond to the
supersymmetric model.

$~$

We thank F. Buccella for discussions.

$~$

\end{document}